\documentclass[
  aps,
  prd,
  reprint,
  amsmath,
  amssymb,
  superscriptaddress,
  floatfix,
  nofootinbib
]{revtex4-2}

\usepackage{graphicx}
\usepackage{bm}
\usepackage{xcolor}
\usepackage{comment}
\usepackage[T1]{fontenc}
\usepackage{hyperref}

\DeclareUnicodeCharacter{2609}{\ensuremath{\odot}}
\DeclareUnicodeCharacter{2500}{--}
\DeclareUnicodeCharacter{03B1}{\ensuremath{\alpha}}
\DeclareUnicodeCharacter{224D}{\ensuremath{\asymp}}

\newcommand{\hi}{H\textsc{i}}
\newcommand{\appref}[1]{\hyperref[#1]{Appendix~\ref*{#1}}}
\newcommand{\secref}[1]{\hyperref[#1]{Section~\ref*{#1}}}

\defcitealias{2026arXiv260701864C}{Chen and Cunnington et~al.}
\newcommand{\citetaliasyear}[1]{\citetalias{#1}~\cite{#1}}

\begin{document}

\title{Emission line stacking on cosmological scales: equivalence to two-point statistics and modelling of the observational effects}

\author{Zhaoting Chen}
\email{zhaoting.chen@roe.ac.uk}
\affiliation{Institute for Astronomy, The University of Edinburgh, Royal Observatory, Edinburgh EH9 3HJ, UK}

\author{Jos\'e Luis Bernal}
\email{jlbernal@ifca.es}
\affiliation{Instituto de F\'isica de Cantabria (IFCA), Universidad de Cantabria--CSIC, Avda.\ de los Castros s/n, 39005 Santander, Spain}

\date{\today}

\begin{abstract}
Emission line stacking measurement for cosmological line intensity mapping (LIM) contains not only the average emission line profile of the stacked sources, but {usually} also the clustering signal{, due to the limited angular resolution of the observations}.
We provide an explicit derivation of the equivalence between the emission line stacking, the two-point cross-correlation function in configuration space {and the power spectrum in Fourier space},
similar to the case between the angular stacking signal for the kinematic Sunyaev-Zel'dovich effect and the angular power spectrum in harmonic space \citep{2025arXiv251214625H,2026arXiv260515947H}.
Using the quadratic estimator formalism, we establish that the emission line stacking signal measurement is equivalent to a quadratic estimator for the cross-correlation function.
The formalism allows for the explicit modelling of the stacked signal,
including the clustering statistics as well as observational effects such as the galaxy selection function and the LIM survey footprint.
In particular, using mock \hi\ intensity mapping data, we show that we can accurately model the effects of foreground removal on the stacked signal.
The correlation function estimator provides significant speed-up for the stacking measurement
and enables modelling of the signal and its covariance for large LIM datasets with MeerKLASS and future SKAO.
\end{abstract}

\maketitle


\section{Introduction}

Stacking is a powerful probe to study the properties and evolution of emission line galaxies.
The baryonic matter within galaxies can be traced by the characteristic emission lines such as the neutral hydrogen (\hi) 21\,cm line
in radio wavelengths and a series of fine-structure and molecular lines in the millimetre and sub-millimetre wavelengths such as [OI] \citep{2023ApJ...953L...3D}, [CII] \citep{2026A&A...710A.309R}, and CO \citep{2026ApJS..282...40P}.
The gas content of galaxies can be used to study the star-formation history of galaxies \citep{2019ARA&A..57..511K} and the relation to the host dark matter halo (see e.g., Refs.~\citep{2020ApJ...902..113I,2024MNRAS.529.4192S}). 
As the brightness of the emission line for a given source scales with the inverse square of the luminosity distance (see, e.g., Ref.~\citep{2017PASA...34...52M}),
sources beyond the local Universe require extreme sensitivity to be detected individually.
By combining the emission line signals from a large number of galaxies,
stacking enables the study of the average properties of the emission line galaxies at high redshifts \citep{2019A&A...631A..78S},
allowing for the study of the cosmic structure formation history.

The stacking technique has also been increasingly used in the study of the large-scale structure of the Universe.
Secondary CMB signals can be detected through stacking the CMB map around optical galaxy positions such as the kinematic Sunyaev-Zel'dovich (kSZ) effect (e.g., Refs.~\citep{2012PhRvL.109d1101H, 2016A&A...586A.140P, 2025PhRvD.112h3509H, 2026arXiv260419744Q, 2026arXiv260419745H})
and thermal Sunyaev-Zel'dovich effect (e.g., Refs.~\citep{2011ApJ...736...39H,2015ApJ...808..151G,2017MNRAS.467.2315V,2019A&A...624A..48D,2022MNRAS.514.1645A}),
probing the baryon distribution in the intergalactic medium.
Stacked weak lensing profiles around clusters shed light on the matter content of the Universe as well as the mass scaling relations
(e.g., Refs.~\citep{2009ApJ...703.2217S,2025A&A...704A.110C,2025A&A...703A..25L}).

In particular, stacking has emerged as a promising tool in line intensity mapping (LIM) surveys.
Intensity mapping surveys map the brightness temperature of the emission line across large areas of the sky blindly,
without resolving individual sources, to achieve sensitivity to the large-scale structure of the Universe
(see, Refs.~\citep{2022A&ARv..30....5B,2026arXiv260203011C} for a review).
As hydrogen gas is the dominant component of the baryonic matter in the Universe,
stacking analysis has been performed extensively in 21\,cm LIM surveys,
using the Parkes telescope \citep{2019MNRAS.489..385T,2020MNRAS.498.5916T},
the CHIME telescope \citep{2023ApJ...947...16A},
and the MeerKAT telescope \citep{2025ApJS..279...19C}.
Efforts have also been made in Lyman-$\alpha$ intensity mapping using HETDEX observations~\citep{2022ApJ...929...90L,2022ApJ...934L..26L},
and in the CO line using COMAP \citep{2024ApJ...965....7D, 2026A&A...707A.256D}
for detecting the cross-correlation stacking signal.
In the future, unprecedented precision will be achieved with the UHF band observation of the MeerKAT Large Area Synoptic Survey (MeerKLASS)~\citep{2016mks..confE..32S,2026arXiv260625955C}
and the \hi\ intensity mapping survey with the mid-frequency array of the Square Kilometre Array Observatory (SKA-Mid)~\citep{2026arXiv260703259W,2026arXiv260630200M} and CHORD~\cite{2026arXiv260709374T}.

As sensitivity and data quality improve,
recent progress has been made in measuring the auto-power spectrum of the \hi\ clustering \citep{2026ApJ..1005L..56P,CHIME:2025cee,2026arXiv261012158M},
indicating that LIM will soon be a competitive probe of cosmology. 
Therefore, it is increasingly important to understand the modelling of the stacked signal,
in order to provide cross-validation and maximise the extraction of cosmological and astrophysical information.
As the angular resolutions of intensity mapping surveys typically range from $\sim$\,arcmin to $\sim$\,deg,
the transverse scales of the measurements are larger than galaxies and even their host dark matter halos.
Therefore, the modelling of the stacked signal goes beyond the average emission line profile of the galaxies,
and requires clustering statistics to be taken into account \citep{2017A&A...607A..89B,2025A&A...702A.247D,2026ApJ..1005....8M}.\footnote{{Optical LIM observations, like those obtained by HETDEX, have better angular resolution and they can target smaller scales in stacking so that the clustering contribution is not included. This is not necessarily the case for SPHEREx, since although its angular resolution is very good, it has poorer spectral resolution, and therefore includes clustering contributions along the line of sight.}}
In turn, the stacked signal can be used to constrain the clustering statistics,
as shown in, e.g., Refs.~\cite{2023ApJ...947...16A,2024MNRAS.535..826R}.
However, the modelling of the signal and its covariance typically relies on forward modelling of the emission line profile of the sources,
which may not be robust against the presence of foregrounds and systematic effects \citep{2025ApJS..279...19C}.
A solid understanding of the statistical properties of the stacked signal is therefore crucial for future data analysis.

The relation between stacking and the clustering two-point correlation function has been extensively discussed in the literature,
in the context of the cosmic infrared background \citep{2006A&A...451..417D,2009ApJ...707.1729M,2013ApJ...779...32V,2015MNRAS.449.4476W,2018MNRAS.480..149H},
weak lensing \citep{2004AJ....127.2544S},
and in the shot-noise dominated regime for LIM surveys \citep{2017MNRAS.470.3220W,2019MNRAS.484.1007W,2021MNRAS.502.5259C}.
Recently, in the context of the kSZ effect, Refs.~\cite{2025arXiv251214625H,2026arXiv260515947H} have shown that the stacked signal
in angular space is equivalent to the two-point angular correlation function (see also Ref.~\citep{2026PhRvD.113h3532S}),
allowing for fast estimation of the signal and its covariance {in harmonic space}, as well as explicit modelling of the signal.

In this paper, we provide a formal derivation of the equivalence between the stacked signal and the two-point correlation function in configuration space,
with a focus on modelling a realistic estimator for the stacked signal.
By utilising the quadratic estimator formalism, we aim to achieve an analytical expression
for the expectation value of the stacking estimator {including the potential signal loss due to foreground cleaning}, and relate it to the modelling of the cross-correlation power spectrum.
Since the cross-correlation power spectrum is one of the major goals of LIM surveys,
the stacking measurement will allow for a direct comparison of the two-point statistics in configuration and Fourier space,
important for science validation and revealing potential systematics.
The equivalence also enables an efficient framework for calculating the covariance of the stacked signal,
ideal for optimising the sensitivity of the stacking measurement and performing model inference.

This paper is organised as follows.
In \secref{sec:formalism}, we provide the mathematical derivation of the equivalence between the stacked signal and the two-point correlation function,
and derive the expectation value of the stacking estimator in the presence of observational effects.
In \secref{sec:validation}, we apply the formalism to mock \hi\ intensity mapping data,
validating the modelling framework numerically.
We discuss the implications of our results and give conclusions in \secref{sec:conclusion}. 
{Finally, we discuss a general estimator for any non-directional stacking in Appendix~\ref{apdx:general}.}

\section{Formalism}
\label{sec:formalism}
We first specify the matrix notation used in this work.
Bold math $\bm{A}$ denotes a vector.
Math blackboard bold $\mathbb{A}$ denotes a matrix.
We do not use Einstein summation convention, so that the summation of indices is explicitly written out with $\sum$.
Subscripts with position vectors or wavenumbers such as  $\bm{A}_{\bm{x}}$ or $\mathbb{A}_{\bm{k}\bm{k}'}$ denote scalars and are specific elements of a vector or matrix.

\subsection{Emission line stacking}
We first write down a generic expression for the emission line stacking,
starting by defining the notation for the emission line signal at different levels.
The original observed intensity map, denoted as $L(\bm{x})\equiv \bm{L}_{\bm{x}}$, goes through a series of operations before being used for stacking.
Since the operations typically include foreground removal,
we denote the map data vector used for stacking as the residual map, $\bm{L}^{\rm res}_{\bm{x}}$.
These operations are \emph{linear}, in the sense that the output map data vector is a linear transformation of the input map data vector,
\begin{equation}
    L_{\rm res}(\bm{x}) \equiv \bm{L}^{\rm res}_{\bm{x}} = \sum_{\bm{x}'} \mathbb{M}_{\bm{x}\bm{x}'} \bm{L}_{\bm{x}'},
\end{equation}
where $\mathbb{M}_{\bm{x}\bm{x}'}$ is the linear transformation matrix that can incorporate the combination of a number of operations.
For example, for PCA foreground cleaning, we have
\begin{equation}
    \mathbb{M}^{\rm PCA}_{\bm{x}\bm{x}'} = \delta^{\rm K}_{\bm{x}\bm{x}'} - \delta^{\rm K}_{\bm{x}_\perp \bm{x}'_\perp}\sum_{n=1}^{N_{\rm fg}} \bm{v}^n_{{x}_\parallel} \bm{v}^n_{{x}_\parallel'},
\label{eq:PCA_matrix}
\end{equation}
where $\delta^{\rm K}_{\bm{x}\bm{x}'}$ is the Kronecker delta function, $\bm{v}^n_{{x}_\parallel}$ is the value of the $n^{\rm th}$ eigenvector at line-of-sight coordinate ${x}_\parallel$.
Or in the case of any smoothing operation, we have
\begin{equation}
    \mathbb{M}^{\rm smooth}_{\bm{x}\bm{x}'} = \mathcal{W}(|\bm{x} - \bm{x}'|),
\end{equation}
where $\mathcal{W}(r)$ is the smoothing window function.
Note that any number of operations can be combined, as long as the order of operations is explicitly defined.
For example, if the foreground cleaning is done after the smoothing, we have $\mathbb{M} = \mathbb{M}^{\rm PCA} \cdot \mathbb{M}^{\rm smooth}$, where $\cdot$ denotes the matrix multiplication.
Other operations can be similarly collapsed into this chain of operations.

After these operations are performed, {we can define the \emph{3D cubelet} around a galaxy at position $\bm{x}_i$ as}
\begin{equation}
    I(\bm{s}|\bm{x}_i) \equiv \bm{I}_{\bm{s}\bm{x}_i} = \sum_{\bm{x}} \mathbb{S}_{\bm{x}_i\bm{x}}^{\bm{s}} \bm{L}^{\rm res}_{\bm{x}},
\end{equation}
where $\mathbb{S}_{\bm{x}_i\bm{x}}^{\bm{s}}$ is the selection function around the galaxy at position $\bm{x}_i$ at separation $\bm{s}$. 
Note that, here $\mathbb{S}^{\bm{s}}$ is a different matrix for each $\bm{s}$.
These cubelets will be later stacked for all galaxies, i.e. by summing over all $\bm{x}_i$, to obtain our stacking measurement.
While $\mathbb{S}$ can be generic, at the 3D cubelet level {to e.g., measure the 3D emission profile}, this is typically a Dirac delta function times a weight,
\begin{equation}
    \mathbb{S}_{\bm{x}_i\bm{x}}^{\bm{s}} = \delta^{\rm K}_{(\bm{x}-\bm{x}_i)\bm{s}} \bm{w}_{\bm{x}},
    \label{eq:S_Dirac}
\end{equation}
where $\bm{w}_{\bm{x}}$ is the weight of the intensity map at position $\bm{x}$, which is typically the inverse noise variance.
We will refrain from any filtering or smoothing that mixes different $\bm{s}$ at this level. 

The \emph{3D stacked signal} is then a linear combination of the 3D cubelets over all galaxies:
\begin{equation}
    I(\bm{s}) \equiv \bm{I}_{\bm{s}} = \frac{1}{Q_0} \sum_i \bm{w}^{\rm gal}_i\, \bm{I}_{\bm{s}\bm{x}_i},
\label{eq:3D_stacking}
\end{equation}
where $Q_0$ is a normalisation factor, and $\bm{w}^{\rm gal}_i$ is the weight of the $i$-th galaxy. For example, $\bm{w}^{\rm gal}_i$ can be a binary weight that selects a particular subsample of the galaxy catalogue.

Here we need to emphasise that, for each cubelet, the weight $\bm{w}^{\rm gal}_i$ is a constant, and is not a function of the position $\bm{s}$.
This is important later when we discuss the stacking as a quadratic estimator. 
Any $\bm{s}$-dependent weighting should be understood as effectively averaging over part of the $\bm{s}$ space,
and therefore is discussed later when we discuss \emph{1D stacking}.
We also do not discuss the case where the 3D stacked signal is averaged into a cylindrical 2D signal,
which is similar to the 1D stacking.

For the case of 1D stacking, we can define the signal as
\begin{equation}
    I_{\rm 1D}(s_\parallel) \equiv \bm{I}^{\rm 1D}_{s_\parallel} = \frac{1}{Q_1} \sum_{\bm{s}'} \bm{S}^{s_\parallel}_{\bm{s}'} \bm{I}_{\bm{s}'},
\label{eq:1D_stacking}
\end{equation}
where $Q_1$ is a normalisation factor, and $\bm{S}^{s_\parallel}_{\bm{s}'}$ is the 1D selection function at separation $s_\parallel$.\footnote{Note that, since the selection function and the average stacking are linear operations that can be implemented in any order, the selection function can be applied at map level using $\mathbb{S}^{\bm{s}}_{\bm{x}_i\bm{x}}$ or at cubelet level using $\bm{S}_{\bm{s}}^{s_\parallel}$, for instance.}
For example, in the simplest case, the average takes place within a certain transverse radius $r_\perp$ around the galaxy so that,
$\bm{S}^{s_\parallel}_{\bm{s}'} = \mathcal{H}\big(r_\perp - |\bm{s}'_\perp|\big)\,\delta^{\rm K}_{{s}'_\parallel {s}_\parallel},$
where $\mathcal{H}(x)$ is the Heaviside step function.
This step function can be generalised to any taper around the centre position similar to techniques of aperture photometry,
for example the so-called ``2D Forced Photometry'' \citep{2026ApJ..1005....8M}.
We can also perform a further averaging of the 1D stacked signal along the spectral direction into the total luminosity, 
e.g. to obtain the average \hi\ mass in radio interferometric emission line observations (see e.g., Ref.~\citep{2022ApJ...935L..13S})
or to maximise the signal-to-noise ratio of the stacked signal \citep{2026ApJ..1005....8M}.
The choice of the 1D selection function is a trade-off between the signal-to-noise ratio and the resolution of the stacked signal.

\subsection{Stacking is two-point statistics: The ideal case}
\label{subsec:ideal_case}
Before discussing the observational effects,
we first establish the mathematical fact that the 3D stacking signal is a two-point correlation function,
and we begin by focusing purely on the signal.

Let us consider the ideal stacking of pure signal. 
In this case, we have $\mathbb{M} = \mathbb{I}$, the identity matrix.
The 3D cubelet selection function is just $\mathbb{S}_{\bm{x}_i\bm{x}}^{\bm{s}} = \delta^{\rm K}_{(\bm{x}-\bm{x}_i)\bm{s}}$.
The 3D stacked cubelet is then
\begin{equation}
    \bm{I}_{\bm{s}\bm{x}_i} = {L}(\bm{x}_i + \bm{s}).
\end{equation}
The cubelets are uniformly weighted, so that $\bm{w}^{\rm gal}_i = 1$.
The normalisation factor is then $Q_0 = N_{\rm gal}$, where $N_{\rm gal}$ is the number of galaxies in the survey volume.

The total stacked signal is then the average of the stacked signal cubelets over all galaxies:
\begin{equation}
\begin{split}
    \bm{I}_{\bm{s}}
    & = \sum_i \bm{I}_{\bm{s}\bm{x}_i} / N_{\rm gal}
    \\ & = \sum_i \int \frac{{\rm d}^3 \bm{x}}{V}
    {L}(\bm{x} + \bm{s})\, \delta_D^3(\bm{x} - \bm{x}_i)
    \Big/ N_{\rm gal}
    \\ & = \frac{1}{\bar{n}_{\rm gal}}
    \int \frac{{\rm d}^3 \bm{x}}{V} {L}(\bm{x} + \bm{s})\, n_{\rm gal}(\bm{x}),
\label{eq:stack_integral}
\end{split}
\end{equation}
where $\bar{n}_{\rm gal} = N_{\rm gal} / V$ is the average galaxy number density in the survey volume.
We have utilised the fact that the galaxy number density field is a sum of the 3D Dirac delta of the galaxy positions
\begin{equation}
    n_{\rm gal}(\bm{x}) = \sum_i \delta_D^3(\bm{x} - \bm{x}_i) / V.
\end{equation}

Under the assumption of homogeneity, the volume integral is equivalent to the ensemble average so that
\begin{equation}
\begin{split}
    I(\bm{s})
    & = \frac{1}{\bar{n}_{\rm gal}}
    \Big\langle L(\bm{x} + \bm{s})\, n_{\rm gal}(\bm{x}) \Big\rangle
    \\ & = \xi \big[L, \delta\big](\bm{s}) + \bar{L},
\label{eq:stack_corr}
\end{split}
\end{equation}
where $\delta = n_{\rm gal} / \bar{n}_{\rm gal} - 1$.
In emission line stacking, the intensity map is always mean-centred so that $\bar{L} = 0$,
either through continuum subtraction or through blind foreground removal.
It is then evident from \autoref{eq:stack_corr} that the stacked signal is the two-point correlation function of the flux density field with the galaxy overdensity field.

\subsection{Stacking is two-point statistics: The observed signal}
\label{subsec:observed_case}
After establishing the intuition in \secref{subsec:ideal_case}, we now consider the observed 3D stacking signal.
The final estimator for the stacked signal, \autoref{eq:3D_stacking}, can be explicitly expanded as
\begin{equation}
    \hat{\bm{I}}_{\bm{s}} = \frac{1}{Q_0} \sum_i \bm{w}^{\rm gal}_i\, \sum_{\bm{x}} \mathbb{S}_{\bm{x}_i\bm{x}}^{\bm{s}} \sum_{\bm{x}'} \mathbb{M}_{\bm{x}\bm{x}'} \bm{L}_{\bm{x}'}.
    \label{eq:general_estimator_configspace}
\end{equation}
We can treat the 3D stacking selection function and the linear transformation matrix as a single operation,
\begin{equation}
    \mathbb{W}_L^{\bm{s}} = \mathbb{S}^{\bm{s}} \cdot \mathbb{M}.
    \label{eq:WLs}
\end{equation}
The 3D stacked signal is then
\begin{equation}
\begin{split}
    \hat{\bm{I}}_{\bm{s}}
    & = \frac{1}{Q_0} \sum_i \bm{w}^{\rm gal}_i\,
    \sum_{\bm{x}} \big[\mathbb{W}_L^{\bm{s}}\big]_{\bm{x}_i\bm{x}} \,\bm{L}_{\bm{x}}
    \\ & = \frac{1}{Q_0} \sum_i \bm{w}^{\rm gal}_i\,
    \big[ \mathbb{W}_L^{\bm{s}} \cdot \bm{L} \big]_{\bm{x}_i}
    \\ & = \frac{V}{Q_0} \int \frac{{\rm d}^3 \bm{x}}{V}
    \big[ \mathbb{W}^{\bm{s}}_L \cdot \bm{L} \big](\bm{x})
    \big[ \bm{w}_{\rm gal}\,\bm{n}_{\rm gal}' \big](\bm{x})
    \\ & = \frac{V}{Q_0} \int \frac{{\rm d}^3 \bm{x}}{V}
    \big[ \mathbb{W}^{\bm{s}}_L \cdot \bm{L} \big](\bm{x})
    \big[ \mathbb{W}_{g} \cdot \bm{n}_{\rm gal} \big](\bm{x}),
\end{split}
\label{eq:3D_stacking_observed_general}
\end{equation}
where $\bm{n}_{\rm gal}'$ {and $\bm{n}_{\rm gal}$ are} the galaxy number density fields of the full observed sample {and the intrinsic number density field without selection weights, respectively}, 
and $\mathbb{W}_{g}$ is a diagonal matrix of the overall weight of the underlying galaxy number density field.
On top of the weights $\bm{w}^{\rm gal}_i$, the total galaxy weight additionally includes the survey geometry and redshift kernel of the observed galaxy sample.

Then the ensemble average for the observed 3D stacked signal is
\begin{equation}
\begin{split}
    \langle \hat{\bm{I}}_{\bm{s}} \rangle
    & = \frac{V}{Q_0} \Big\langle
    \int \frac{{\rm d}^3 \bm{x}}{V}
    \big[ \mathbb{W}_L^{\bm{s}} \cdot \bm{L} \big](\bm{x})
    \big[\mathbb{W}_g \cdot \bm{n}_{\rm gal} \big](\bm{x})
    \Big\rangle
    \\ & = \frac{V}{Q_0}\,
    \xi \big[\mathbb{W}_L^{\bm{s}} \cdot \bm{L},
    \mathbb{W}_g \cdot \bm{n}_{\rm gal}\big](\bm{0}),
\end{split}
\label{eq:3D_stacking_observed_WLs}
\end{equation}
{which corresponds to the correlation function, apparently evaluated at zero lag. However, note that the scale information in the two-point correlation that the stacking signal entails is encoded in the matrix $\mathbb{W}_L^{\bm{s}}$; we show below this dependence explicitly.}

To further simplify \autoref{eq:3D_stacking_observed_general} and \autoref{eq:3D_stacking_observed_WLs}, we need to consider a special class of $\mathbb{W}_L^{\bm{s}}$. 
{We can always separate the selection function from the filtering or smoothing of the map (see \autoref{eq:WLs}). 
For the case in which $\mathbb{S}^{\bm{s}}$ is proportional to a Dirac delta function we have}
\begin{equation}
    \big[\mathbb{W}_L^{\bm{s}} \big]_{\bm{x}_i\bm{x}'}  
    \propto\sum_{\bm{x}} \delta^{\rm K}_{(\bm{x}-\bm{x}_i)\bm{s}} \big[ \mathbb{W}^L \big]_{\bm{x}\bm{x}'}.
\end{equation}
Note that this is usually the case for emission line stacking.
Here the Dirac delta function requirement is simply indicating that
the measured signal is at the level of the 3D stacked cubelets. For example, if the intensity map voxel is weighted by the hit counts $\bm{h}_{\bm{x}}$, and then 
selected at separation $\bm{s}$ for each galaxy position $\bm{x}_i$, we have
\begin{equation}
    \mathbb{W}^L_{\bm{x}\bm{x}'} = \bm{h}_{\bm{x}}\, \mathbb{M}_{\bm{x}\bm{x}'}\,.
\end{equation} 

For this case, the ensemble average for the observed 3D stacked signal is
\begin{equation}
\begin{split}
    \langle \hat{\bm{I}}_{\bm{s}} \rangle
    & = \frac{V}{Q_0} \Big\langle
    \int \frac{{\rm d}^3 \bm{x}}{V}
    \big[ \mathbb{W}_L \cdot \bm{L} \big](\bm{x} + \bm{s})
    \big[\mathbb{W}_g \cdot \bm{n}_{\rm gal} \big](\bm{x})
    \Big\rangle
    \\ & = \frac{V}{Q_0}\,
    \xi \big[
      \mathbb{W}_L \cdot \bm{L},
      \mathbb{W}_g \cdot \bm{n}_{\rm gal}
    \big](\bm{s})\,,
\end{split}
\label{eq:3D_stacking_observed}
\end{equation}
As one can see {in these equations}, the observed 3D stacking signal is the two-point correlation function between the ``observed'' flux density field and the ``observed'' galaxy number density field at separation $\bm{s}$.
We define the ``observed'' flux density field as
\begin{equation}
    L_{\rm obs}(\bm{x}) = \big[\mathbb{W}_L \cdot \bm{L}\big](\bm{x})
\end{equation}
and the ``observed'' galaxy number density field as
\begin{equation}
    n_{\rm obs}(\bm{x}) = \big[\mathbb{W}_g \cdot \bm{n}_{\rm gal}\big](\bm{x})\,.
\end{equation}

\subsection{Stacking as a quadratic estimator}
\label{subsec:quadratic_general}
\subsubsection{Estimating the stacking signal from Fourier space measurements}
We have established that the stacking signal is, in fact, a two-point correlation function.
Therefore, we can fully utilise the machinery of the quadratic estimator to model the observed stacking signal.
It will naturally give rise to the modelling of the stacking signal from the clustering statistics,
as well as the appropriate normalisation and window function to correct for various observational effects,
such as the galaxy selection function and foreground removal effects.

We first write the Fourier convention in a discrete form, as
\begin{equation}
    \tilde{\bm{f}}_{\bm{k}} = \sum_{\bm{x}} \mathbb{F}_{\bm{k}\bm{x}}\bm{f}_{\bm{x}},\, \qquad \qquad \bm{f}_{\bm{x}} = \sum_{\bm{k}} \mathbb{F}^{-1}_{\bm{x}\bm{k}}\tilde{\bm{f}}_{\bm{k}},
\label{eq:fourier_convention_dft}
\end{equation}
where $\mathbb{F}$ and $\mathbb{F}^{-1}$ are the forward and inverse discrete Fourier transform (DFT) matrices, respectively {and we denote Fourier-space quantities with a tilde}.
The Fourier convention typically used in cosmology implies that,
\begin{equation}
    \mathbb{F}_{\bm{k}\bm{x}} = \frac{1}{N_{\rm grid}} {\rm exp}\big[{-i\bm{k}\cdot\bm{x}}\big],\qquad
    \mathbb{F}^{-1} = {N_{\rm grid}}\, \mathbb{F}^\dagger,
\label{eq:f_inverse}
\end{equation}
where $N_{\rm grid}$ is the number of grids in the DFT box, and $\dagger$ denotes the conjugate transpose.

We can connect the standard stacked signal with the power spectrum noting that
\begin{equation}
\begin{split}
    I(\bm{s})
    & = \xi \big[L, \delta\big](\bm{s}) = \frac{1}{V}\,\mathcal{F}^{-1}\Big[\mathcal{P}[L, \delta]  \Big]
    \\ & = \frac{1}{V}\,\sum_{\bm{k}}\mathbb{F}^{-1}_{\bm{s}\bm{k}}\, \bm{P}_{\bm{k}},
\end{split}
\end{equation}
where $\mathcal{P}[...,...]$ denotes the cross-power spectrum between two fields, $\mathcal{F}$ denotes the Fourier transform operator, and $\bm{P}_{\bm{k}} = P(\bm{k})$ is the cross-power spectrum between {the intrinsic, unweighted} $\delta\equiv n_{\rm gal}/\bar{n}_{\rm gal}$ and $L$.

Similar to \autoref{eq:3D_stacking_observed}, we can rewrite the stacked signal as
\begin{equation}
\begin{split}
    \langle \hat{\bm{I}}_{\bm{s}} \rangle
    & = \frac{V\,\bar{n}_{\rm gal}}{Q_0 \bar{n}_{\rm gal}}\,
    \xi \big[\bm{L}_{\rm obs}, \bm{n}_{\rm obs}\big](\bm{s})
    \\ & = \frac{V\,\bar{n}_{\rm gal}}{Q_0}\,
    \xi \Big[
      \bm{L}_{\rm obs},
      \mathbb{W}_g \cdot \frac{\bm{n}_{\rm gal}}{\bar{n}_{\rm gal}}
    \Big](\bm{s})
    \\ & = \frac{\bar{n}_{\rm gal}}{Q_0}\,
    \mathbb{F}^{-1}\Big[
      \mathcal{P}[\bm{L}_{\rm obs}, \mathbb{W}_g \cdot \bm{\delta}](\bm{k})
    \Big](\bm{s})
    \\ & = \frac{1}{Q_0'}\,
    \mathbb{F}^{-1}\Big[
      \mathcal{P}[\bm{L}_{\rm obs}, \bm{\delta}_{\rm obs}](\bm{k})
    \Big](\bm{s}),
\end{split}
\label{eq:stack_estimator}
\end{equation}
where we denote $Q_0' = Q_0 / \bar{n}_{\rm gal}$ and $\bm{\delta}_{\rm obs} = \mathbb{W}_g \cdot \bm{\delta}$.
Note that the division by $\bar{n}_{\rm gal}$ and multiplication by $\mathbb{W}_g$ are commutative because $\mathbb{W}_g$ is a diagonal matrix.

\subsubsection{Quadratic estimator for stacking without foreground removal}
We first consider a special case. Suppose there is no foreground removal, filtering etc.,
so that $L_{\rm obs}(\bm{x}) =  w_L(\bm{x}) L(\bm{x})$.
We can write down the normalisation in \autoref{eq:stack_estimator} as the sum of the map weights multiplied by the galaxy weights over the selected galaxies,
\begin{equation}
    Q_0 = \sum_i \bm{w}^{L}_{\bm{x}_i} \bm{w}^{\rm gal}_{\bm{x}_i}.
\end{equation}
The fact that we sum over $w_{L}(\bm{x}_i)$ instead of $w_{L}(\bm{x}_i + \bm{s})$ will be evident
if we consider that,
\begin{equation}
\begin{split}
    \sum_i \bm{w}^{L}_{\bm{x}_i} \bm{w}^{\rm gal}_{\bm{x}_i} & =V \int \frac{{\rm d}^3 \bm{x}}{V} w_{L}(\bm{x})w_{g}(\bm{x}) \, n_{\rm obs}(\bm{x})
    \\ & = V \int \frac{{\rm d}^3 \bm{x}}{V} w_{L}(\bm{x}) \, W_{g}(\bm{x}) \, n_{\rm gal}(\bm{x})
    \\ & = V\,\langle w_{L} \, W_{g} \, n_{\rm gal} \rangle 
    \\ & = V\,\langle w_{L} \, W_{g} \rangle \,\bar{n}_{\rm gal}\,,
\end{split}
\end{equation}
where $ W_{g}n_{\rm gal} = w_{g} n_{\rm obs}$, 
as we have defined in \autoref{eq:3D_stacking_observed_general},
and we used the fact that the galaxy number density field is not correlated with the weights.
Note that the above equation no longer holds if we instead sum over $w^{L}_{\bm{x}_i + \bm{s}}$.
The overall normalisation factor is then given by,
\begin{equation}
    Q_0' = V\,\langle w_{L}\, w_{g} \rangle.
\end{equation}
We start from a simple normalisation over the sum of the weights and recover the well-known
normalisation factor for the power spectrum quadratic estimator $\langle w_1 w_2\rangle$ (see, e.g., Ref.~\citep{2019MNRAS.489..153B}).

Now we turn to the normalisation factor for any generic stacking procedure.\footnote{{We neglect any averaging or tapering in the selection function here for simplicity, since we can always consider it after stacking the 3D cubelets.}}
We start from a generic power spectrum estimator,
\begin{equation}
\begin{split}
    \hat{P}
    & = \mathcal{P}\Big[
      \sum_{\bm{x}_1'}\mathbb{W}^L_{\bm{x}_1\bm{x}_1'} \, \bm{L}_{\bm{x}_1'},
      \sum_{\bm{x}_2'}\mathbb{W}^g_{\bm{x}_2\bm{x}_2'} \, \bm{\delta}_{\bm{x}_2'}
    \Big].
\end{split}
\end{equation}
The estimator can be expanded as,
\begin{equation}
\begin{split}
    \hat{\bm{P}}_{\bm{k}}
    & = V \,\Big|
      \mathbb{F}[\bm{W}_L\cdot \bm{L}](\bm{k})\,
      \mathbb{F}[\mathbb{W}_g\cdot \bm{\delta}]^*(\bm{k})
    \Big|_{\rm Re}
    \\ & = V\sum_{\substack{\bm{x}_1'\bm{x}_2'\\\bm{x}_1\bm{x}_2}}
    \Big|
      \mathbb{F}_{\bm{k}\bm{x}_1} \, \mathbb{W}^L_{\bm{x}_1\bm{x}_1'} \, \bm{L}_{\bm{x}_1'}
      \mathbb{F}_{\bm{k}\bm{x}_2}^* \, \mathbb{W}^g_{\bm{x}_2\bm{x}_2'} \, \bm{\delta}_{\bm{x}_2'}
    \Big|_{\rm Re}
    \\ & = V\sum_{\substack{\bm{x}_1'\bm{x}_2'\\\bm{x}_1\bm{x}_2\\\bm{k}_1\bm{k}_2}}
    \Big|
      \mathbb{F}_{\bm{k}\bm{x}_1} \, \mathbb{W}^L_{\bm{x}_1\bm{x}_1'}
      \, \mathbb{F}^{-1}_{\bm{x}_1'\bm{k}_1} \, \tilde{\bm{L}}_{\bm{k}_1}
    \\ & \qquad\qquad \times
      \mathbb{F}_{\bm{k}\bm{x}_2}^* \, \mathbb{W}^g_{\bm{x}_2\bm{x}_2'}
      \, \big(\mathbb{F}^{-1}_{\bm{x}_2'\bm{k}_2}\big)^* \, \tilde{\bm{\delta}}_{\bm{k}_2}^*
    \Big|_{\rm Re}.
\end{split}
\label{eq:psest_general}
\end{equation}
For clarity, we can define the Fourier transform of a matrix as follows,
\begin{equation}
    \tilde{\mathbb{W}}_{\bm{k}_1\bm{k}_2} = \sum_{\bm{x}_1\bm{x}_2}\mathbb{F}_{\bm{k}_1\bm{x}_1} \, \mathbb{W}_{\bm{x}_1\bm{x}_2} \, \mathbb{F}_{\bm{x}_2\bm{k}_2}^{-1},
\end{equation}
so that the estimator in \autoref{eq:psest_general} can be written as,
\begin{equation}
\begin{split}
    \hat{\bm{P}}_{\bm{k}}
    & = V \sum_{\bm{k}_1\bm{k}_2}\Big|
      \tilde{\mathbb{W}}^L_{\bm{k}\bm{k}_1}\,\tilde{\bm{L}}_{\bm{k}_1}
      \big(\tilde{\mathbb{W}}^\dagger_g\big)_{\bm{k}_2\bm{k}}\, \tilde{\bm{\delta}}^*_{\bm{k}_2}
    \Big|_{\rm Re}.
\end{split}
\end{equation}

The ensemble average of the estimator is then given by,
\begin{equation}
\begin{split}
    \langle \hat{\bm{P}}_{\bm{k}} \rangle
    & = V \sum_{\bm{k}_1\bm{k}_2}\Big|
      \tilde{\mathbb{W}}^L_{\bm{k}\bm{k}_1}
      \, \big(\tilde{\mathbb{W}}_g^{\dagger}\big)_{\bm{k}_2\bm{k}}
      \big\langle \tilde{\bm{L}}_{\bm{k}_1} \tilde{\bm{\delta}}^*_{\bm{k}_2} \big\rangle
    \Big|_{\rm Re}
    \\ & = \sum_{\bm{k}_1\bm{k}_2}
    \tilde{\mathbb{W}}^L_{\bm{k}\bm{k}_1}
    \, \big(\tilde{\mathbb{W}}_g^{\dagger}\big)_{\bm{k}_2\bm{k}}
    \, P_{\bm{k}_1} \, \delta^{\rm K}_{\bm{k}_1\bm{k}_2}
    \\ & = \sum_{\bm{k}'}
    \Big(
      \tilde{\mathbb{W}}^L_{\bm{k}\bm{k}'}
      \big(\tilde{\mathbb{W}}_g^\dagger\big)_{\bm{k}'\bm{k}}
    \Big) P_{\bm{k}'}.
\end{split}
\end{equation}
We therefore obtain the window function for the power spectrum estimator as,
\begin{equation}
    \mathbb{H}_{\bm{k}\bm{k}'} = \tilde{\mathbb{W}}^L_{\bm{k}\bm{k}'} \big(\tilde{\mathbb{W}}_g^\dagger\big)_{\bm{k}'\bm{k}}.
\label{eq:h_general}
\end{equation}
The associated {k-dependent} normalisation factor {of the power spectrum estimator that will appear in the estimator of the stacking signal (\autoref{eq:stack_estimator})} is therefore,
\begin{equation}
    \bar{Q}_{\bm{k}} = \sum_{\bm{k}'}\mathbb{H}_{\bm{k}\bm{k}'} = \big(  \tilde{\mathbb{W}}_L \cdot \tilde{\mathbb{W}}_g^\dagger \big)_{\bm{k}\bm{k}},
\label{eq:q0_general}
\end{equation}
so that in the limit of $P_{\bm{k}'}$ being a constant $P_0$ over the non-zero part of $\mathbb{H}_{\bm{k}\bm{k}'}$, $\langle \hat{P}_{\bm{k}} \rangle = \bar{Q} P_0$. 

For pedagogical purposes, let us re-derive the normalisation factor in the case of $\mathbb{W}_L$ and $\mathbb{W}_g$ being diagonal matrices.
In this case, the operators are simply
\begin{equation}
\begin{split}
    \mathbb{W}^L_{\bm{x}\bm{x}'} &= \delta^{\rm K}_{\bm{x}\bm{x}'}\,\bm{w}^{L}_{\bm{x}},
    \\ \mathbb{W}^g_{\bm{x}\bm{x}'} &= \delta^{\rm K}_{\bm{x}\bm{x}'}\,\bm{W}^{g}_{\bm{x}}.
\end{split}
\end{equation}
In this case, the normalisation factor is
\begin{equation}
\begin{split}
    \bar{Q}_{\bm{k}} & = \big(  \tilde{\mathbb{W}}_L \cdot \tilde{\mathbb{W}}_g^\dagger \big)_{\bm{k}\bm{k}} = \Big(\mathbb{F}\cdot \mathbb{W}_L \cdot \mathbb{W}_g^{\rm T} \cdot \mathbb{F}^{-1}\Big)_{\bm{k}\bm{k}},
    \\ & = \sum_{\bm{x}} \mathbb{F}_{\bm{k}\bm{x}} \, \bm{w}^{L}_{\bm{x}} \, \bm{W}^{g}_{\bm{x}} \, \mathbb{F}^{-1}_{\bm{x}\bm{k}}
    \\ & = \sum_{\bm{x}} \frac{\bm{w}^{L}_{\bm{x}} \, \bm{W}^{g}_{\bm{x}} \, {\rm exp}\big[{-i\bm{k}\cdot\bm{x}}\big] \, {\rm exp}\big[{i\bm{k}\cdot\bm{x}}\big]}{N_{\rm grid}}
    \\ & = \big\langle {w}_{L}\, {W}_{g} \big\rangle,
\end{split}
\end{equation}
which in the expression for the stacking signal cancels with $Q_0'$ and gives the correct result $\langle \hat{I}_{\bm{s}}\rangle = \mathbb{F}^{-1}[P_{\bm{k}}/V]$. In the above equation, we utilised the Fourier convention in \autoref{eq:f_inverse}. 

\subsubsection{Quadratic estimator for stacking accounting for foreground removal}
Now we turn to the case where foreground removal is applied.
While \autoref{eq:q0_general} is valid for any operator, {3D matrices are expensive to compute. In any case, we are interested in specific cases of intensity mapping stacked around galaxies, so that} we can simplify the derivation for the case where only PCA along the frequency axis is applied. 
The foreground cleaning operator is then given by \autoref{eq:PCA_matrix}.
{In this case, the weighting matrix for the galaxy number density field is diagonal and the line-intensity maps are modified by a diagonal cleaning matrix and a foreground cleaning matrix given, respectively, by}
\begin{equation}
    \mathbb{W}_g = {\rm diag}\big[\bm{W}_g\big]\,,\qquad \mathbb{W}_L = {\rm diag}\big[\bm{w}_L \big]\cdot
    \mathbb{M}_{\rm PCA}.
\end{equation}
{The Fourier transforms of the two matrices are then given by}
\begin{equation}
    \tilde{\mathbb{W}}^g_{\bm{k}\bm{k}'} = \tilde{\bm{W}}^g_{\bm{k}-\bm{k}'}\,,
\end{equation}
where $\tilde{\bm{W}}^g_{\bm{k}-\bm{k}'}$ denotes the convolution in $\bm{k}$-space, and
\begin{equation}
\begin{split}
    \tilde{\mathbb{W}}^L_{\bm{k}\bm{k}'}
    & = \sum_{\bm{q}}\tilde{\bm{w}}^L_{\bm{k}-\bm{q}}\,
    \Big[
      \delta^{\rm K}_{\bm{q}\bm{k}'}
      - \delta^{\rm K}_{\bm{q}_\perp\bm{k}'_\perp}
      \sum_i \tilde{\bm{v}}^i_{q_\parallel}
      \big(\tilde{\bm{v}}^i_{k'_\parallel}\big)^*
    \Big]
    \\ & = \tilde{\bm{w}}^L_{\bm{k}-\bm{k}'}
-\sum_{q_\parallel}
\tilde{\bm{w}}^L_{\bm{k}_\perp - \bm{k}'_\perp,k_\parallel - q_\parallel}
    \sum_{i} \tilde{\bm{v}}^i_{q_\parallel}
    \big(\tilde{\bm{v}}^i_{k'_\parallel}\big)^*.
\end{split}
\end{equation}
{The normalisation factor is then}
\begin{equation}
\begin{split}
    & \sum_{\bm{k}'}\tilde{\mathbb{W}}^L_{\bm{k}\bm{k}'}
    \big(\tilde{\mathbb{W}}_g^\dagger\big)_{\bm{k}'\bm{k}}
    \\ & \quad =
    \sum_{\bm{k}'}\tilde{\bm{W}}^g_{\bm{k}-\bm{k}'}\tilde{\bm{w}}^L_{\bm{k}-\bm{k}'} 
    \\ & \quad \quad \, -
    \sum_{\bm{k}',q_\parallel} \tilde{\bm{W}}^g_{\bm{k}-\bm{k}'}
    \tilde{\bm{w}}^L_{\bm{k}_\perp-\bm{k}'_\perp,k_\parallel - q_\parallel}
    \sum_{n} \tilde{\bm{v}}^n_{q_\parallel}
    \big(\tilde{\bm{v}}^n_{k'_\parallel}\big)^*.
\end{split}
\end{equation}
We can simplify this expression above noting that, in configuration space, the intensity map weighting is typically smooth along the spectral direction. 
In the limit of no variation of the weights along the line of sight, their expression in Fourier space becomes
\begin{equation}
    \tilde{w}_L(\bm{k}_\perp-\bm{k}'_\perp,k_\parallel - q_\parallel) \approx \tilde{w}_L(\bm{k}-\bm{k}')\,\delta^{\rm K}_{(k_\parallel - q_\parallel)\,0},
\end{equation}
or in vector notation, $\tilde{\bm{w}}^L_{\bm{k}_\perp-\bm{k}'_\perp,k_\parallel - q_\parallel} \approx \tilde{\bm{w}}^L_{\bm{k}-\bm{k}'}\,\delta^{\rm K}_{(k_\parallel - q_\parallel)\,0}$.
{In most situations, intensity mapping observations are in this limit. When sampling instead varies across frequency channels, the smoothness can be enforced by applying a frequency taper as part of the weighting $w_L$, a standard practice in data analysis, so that the approximation above is valid. Applying this approximation, the normalisation factor becomes}
\begin{equation}
\begin{split}
    & \sum_{\bm{k}'}
    \tilde{\mathbb{W}}^L_{\bm{k}\bm{k}'}
    \big(\tilde{\mathbb{W}}_g^\dagger\big)_{\bm{k}'\bm{k}}
    \\ & \quad \approx
    \sum_{\bm{k}'}\tilde{\bm{W}}^g_{\bm{k}-\bm{k}'}\tilde{\bm{w}}^L_{\bm{k}-\bm{k}'}
    \Big[1 - \sum_{i} |\tilde{\bm{v}}^n_{k_\parallel}|^2\Big]
    \\ & \quad = \langle {w}_L {W}_g \rangle\, \mathcal{T}(k_\parallel)\,,
\end{split}
\end{equation}
{where we have applied Parseval's theorem to obtain the result, and we have defined the transfer function as}~\citep{2025MNRAS.542L...1C}
\begin{equation}
    \mathcal{T}(k_\parallel) = 1 - \sum_{i} |\tilde{\bm{v}}^n_{k_\parallel}|^2.
\end{equation}

Following these derivations, the expectation value of the stacking estimator can be approximated as
\begin{equation}
    \langle \hat{\bm{I}}_{\bm{s}} \rangle = \frac{1}{{Q}_0'}\sum_{\bm{k}}\mathbb{F}^{-1}_{\bm{s}\bm{k}}\, P_{\bm{k}} \, \bar{Q}_{\bm{k}} \, \mathcal{T}({k}_\parallel) = \frac{1}{V}\Big(\mathcal{F}^{-1}\big[\mathcal{T}\,P\big]\Big)(\bm{s})\,.
\label{eq:stack_3d_with_pca}
\end{equation}
Then, the stacked signal is the inverse Fourier transform of the signal-loss-corrected power spectrum.

Our formalism then points to the same conclusion found in the literature such as Refs.~\cite{2025arXiv251214625H,2026arXiv260515947H}.
The stacked signal can be obtained by simply applying the inverse Fourier transform of the quadratic power spectrum estimator,
\begin{equation}
    \hat{\bm{I}}_{\bm{s}} = \frac{1}{V}\,\mathbb{F}^{-1}\Big[\hat{\bm{P}}\Big](\bm{s}),
\label{eq:stack_is_ift_ps}
\end{equation}
as long as the effective weightings of the intensity maps and the galaxy number density field are consistent. {In this work we have generalised this statement to any potential selection function and, importantly, have explicitly shown how to compute the estimator tracking all the weights and (linear) operations applied to the observed maps.}

In most cases, the stacking estimator requires much more computational time because it loops over each galaxy position.
The power spectrum estimator, on the other hand, can be significantly sped up
by gridding the galaxy positions onto the grid of the intensity maps
and then applying the estimator at the DFT grid level.
In the case of a large number of galaxies, the latter approach is clearly much more efficient, {as shown for the case of the kSZ effect in Ref.~\citep{2025arXiv251214625H}}.

\subsection{From 3D to 1D stacking: the transverse-plane projection}
\label{subsec:projection}

We have established the equivalence between the 3D stacked signal and the correlation function estimator.
In practice, however, the 3D stacked cubelets are often integrated into a lower-dimensional signal,
for example by averaging over the transverse plane into a 1D stacked spectrum {or radial emission profiles}, or directly {applying an averaging selection function to transform the cubelet into a scalar, as in the case of the integrated flux}
(see e.g., Ref.~\citep{2026ApJ..1005....8M}).
For simplicity, we focus on the 1D stacking estimator, as defined in \autoref{eq:1D_stacking}, and note that the derivations
can be extended to {any other selection function applied} as shown in \appref{apdx:general}.

The 1D selection function in \autoref{eq:1D_stacking} is an aperture of transverse radius $r_\perp$ around the source position,
\begin{equation}
    \bm{S}^{s_\parallel}_{\bm{s}'} = W_\perp(\bm{s}'_\perp)\,\delta^{\rm K}_{s'_\parallel s_\parallel},
    \label{eq:1d_selection}
\end{equation}
where $W_\perp(\bm{s}_\perp)$ is the transverse aperture window function, which for a circular top-hat aperture is $\mathcal{H}\big(r_\perp - |\bm{s}_\perp|\big)$.

The {additional} normalisation factor in \autoref{eq:1D_stacking} {to go from 3 to 1 dimensions} is then naturally the sum of the window function over the transverse plane,
\begin{equation}
    Q_1 = \sum_{\bm{s}_\perp} W_\perp(\bm{s}_\perp)\,,
    \label{eq:aperture_area}
\end{equation}
which in the continuous limit becomes the aperture area. 

Taking the expectation value of \autoref{eq:1D_stacking} and using that the 3D stacked signal is the two-point correlation function,
\begin{equation}
\begin{split}
    \langle \hat{\bm{I}}^{\rm 1D}_{s_\parallel} \rangle
    & = \frac{1}{Q_1} \sum_{\bm{s}_\perp} W_\perp(\bm{s}_\perp)\,\langle \hat{\bm{I}}_{\bm{s}_\perp, s_\parallel} \rangle
    \\ & = \frac{1}{Q_1} \sum_{\bm{s}_\perp} W_\perp(\bm{s}_\perp)\,\langle \hat{\xi}\rangle(\bm{s}_\perp, s_\parallel) 
    \\ & = 
    \frac{1}{V Q_1} \sum_{\bm{s}_\perp} W_\perp(\bm{s}_\perp)
    \\ & \quad \times \sum_{\bm{k}} {\rm e}^{i\bm{k}_\perp\cdot\bm{s}_\perp}\, {\rm e}^{i k_\parallel s_\parallel}\, \langle \hat{P}\rangle(\bm{k}_\perp, k_\parallel)
    \\ & = \frac{1}{Q_1}\sum_{k_\parallel} {\rm e}^{i k_\parallel s_\parallel}\, P_{\rm 1D}(k_\parallel),
\end{split}
\label{eq:1d_power}
\end{equation}
where we have defined the effective 1D power spectrum as the aperture-weighted average of the 3D power spectrum over the transverse plane, {given in a discrete grid as}
\begin{equation}
    P_{\rm 1D}(k_\parallel) = \frac{1}{V N_\perp} \sum_{\bm{k}_\perp} \tilde{W}_\perp(\bm{k}_\perp)\, P(\bm{k}_\perp, k_\parallel),
    \label{eq:p1d}
\end{equation}
where the sum runs over the $N_\perp$ transverse modes of the discrete Fourier grid of the cubelet, $|\bm{k}_\perp| \le k_{\rm Nyq} \equiv \pi/\Delta s_\perp$, with $\Delta s_\perp$ the transverse voxel size,
and $\tilde{W}_\perp(\bm{k}_\perp)$ is the normalised discrete Fourier transform of the aperture window,
\begin{equation}
    \tilde{W}_\perp(\bm{k}_\perp) = \frac{1}{Q_1} \sum_{\bm{s}_\perp} W_\perp(\bm{s}_\perp)\, {\rm e}^{-i\bm{k}_\perp\cdot\bm{s}_\perp},
    \label{eq:aperture_window}
\end{equation}
with the normalisation $\tilde{W}_\perp(\bm{0}) = 1$.
\autoref{eq:1d_power} gives the standard form of the 1D power spectrum in the literature,
in particular for Lyman-$\alpha$ forest measurements (see, e.g., Eq.~6 of Ref.~\cite{2026arXiv260727413C}).

In the shot-noise limit, the aperture is shrunk to a single voxel containing the point source,
i.e. the aperture window function is a Kronecker delta, $W_\perp(\bm{s}_\perp) = \delta^{\rm K}_{\bm{s}_\perp\,\bm{0}}$,
so that $Q_1 = 1$ and $\tilde{W}_\perp(\bm{k}_\perp) = 1$ for all $\bm{k}_\perp$.
For a cubelet with a transverse size $R_\perp$, the smallest transverse wavenumber is $\pi/R_\perp$.
For large $\pi/R_\perp$, where the signal is already dominated by the shot noise, \autoref{eq:p1d} then gives
\begin{equation}
\begin{split}
     P_{\rm 1D}(k_\parallel)\big|_{\substack{\bm{s}_\perp \to 0 \\ \pi/R_\perp \to \infty}}
    & = \frac{1}{N_\perp}
    \sum_{\bm{k}_\perp} P(|\bm{k}_\perp| = \infty, k_\parallel) 
\\ &     = P_{\rm SN}(k_\parallel),
\end{split}
    \label{eq:p_sn_limit}
\end{equation}
where $P_{\rm SN}(k_\parallel)$ is the cross shot-noise power spectrum.

We emphasise that, in this paper, we consider the deterministic shot noise arising from the point source nature of the galaxies.
This is in contrast to the stochastic shot noise arising from the stochastic distribution of the galaxy overdensity field in relation
to the underlying matter distribution (see Ref.~\cite{2023PhRvD.108h3528O} for a detailed discussion).
We also note that the shot noise in LIM is not a constant,
as the point-source nature of the emission is broadened by the internal velocity dispersion of the sources along the line of sight (see e.g., Ref.~\citep{2020ApJ...895...34Z}).

To make the connection to the shot noise explicit, we first consider sources without internal velocity dispersion.
The 1D stacked signal is then
\begin{equation}
    I(s_\parallel) = \frac{\sum_i \bm{w}_i \bm{L}_i}{\sum_i \bm{w}_i} \delta^{\rm K}_{({s}_\parallel)0},
\end{equation}
where $\bm{L}_i$ is the line emission of the $i$-th tracer.
We can further Fourier transform the stacked signal and rewrite it as
\begin{equation}
\begin{split}
    P_{\rm SN}^0 & = V\frac{\sum_i \bm{w}_i \bm{L}_i}{\sum_i \bm{w}_i} = C_L \bar{M}_L,
\end{split}
\label{eq:p_sn_0}
\end{equation}
where $\bar{M}_L$ is the weighted average line emission mass (e.g. the \hi\ mass) of the tracer given by
\begin{equation}
    \bar{M}_L = \frac{\sum_i \bm{w}_i \bm{M}_i}{\sum_i \bm{w}_i},
\end{equation}
and $C_L$ is the conversion factor from the tracer emission line mass density (such as the \hi\ mass) to the flux density.
This gives the familiar shot noise power spectrum in Ref.~\cite{2017MNRAS.470.3220W}.

\subsection{Covariance estimation}
\label{subsec:covariance}
We now briefly discuss the implications of the results for covariance estimation.
The equivalence between the stacking estimator and the correlation function allows us to
estimate the covariance of the stacked signal directly from the covariance estimation of the power spectrum.
In a large scale structure survey, the power spectrum covariance can be accurately estimated in various ways,
for example through analytical modelling (see e.g., Ref.~\citep{2025JCAP...01..145R}) or numerical simulations (see e.g., Ref.~\citep{2025JCAP...04..055F}).
Therefore, suppose we have obtained the power spectrum covariance,
\begin{equation}
    {\mathbb{C}}_{\hat{\bm{P}}} = \langle \hat{\bm{P}} \hat{\bm{P}}^\dagger \rangle - \langle \hat{\bm{P}} \rangle \langle \hat{\bm{P}}^\dagger \rangle,
\end{equation}
we can then estimate the covariance of the 3D stacked signal following \autoref{eq:stack_is_ift_ps},
\begin{equation}
    {\mathbb{C}}_{\hat{\bm{I}}} = \frac{1}{V^2}\,\mathbb{F}^{-1} \cdot {\mathbb{C}}_{\hat{\bm{P}}} \cdot \mathbb{F}.
\end{equation}
Then, following \autoref{eq:1D_stacking}, we can estimate the covariance of the 1D stacked signal as
\begin{equation}
\begin{split}
    {\mathbb{C}}_{\hat{\bm{I}}_{\rm 1D}}
    & = \langle \hat{\bm{I}}_{\rm 1D} \hat{\bm{I}}_{\rm 1D}^\dagger \rangle
    - \langle \hat{\bm{I}}_{\rm 1D} \rangle
    \langle \hat{\bm{I}}_{\rm 1D}^\dagger \rangle
    \\ & = \frac{1}{V^2}\, \mathbb{F}_\parallel^{-1} \Big[
      \langle \hat{\bm{P}}_{\rm 1D} \hat{\bm{P}}_{\rm 1D}^\dagger \rangle
      - \langle \hat{\bm{P}}_{\rm 1D} \rangle
      \langle \hat{\bm{P}}_{\rm 1D}^\dagger \rangle
    \Big] \mathbb{F}_\parallel
    \\ & = \mathbb{C}_{\hat{\bm{\xi}}_{\rm 1D}},
\end{split}
\label{eq:covariance_1d}
\end{equation}
where the 1D correlation function estimator $\hat{\bm{\xi}}_{\rm 1D}$ is the inverse Fourier transform of the 1D power spectrum estimator.

\section{Numerical Validation}
\label{sec:validation}
\begin{figure*}
    \includegraphics[width=\textwidth]{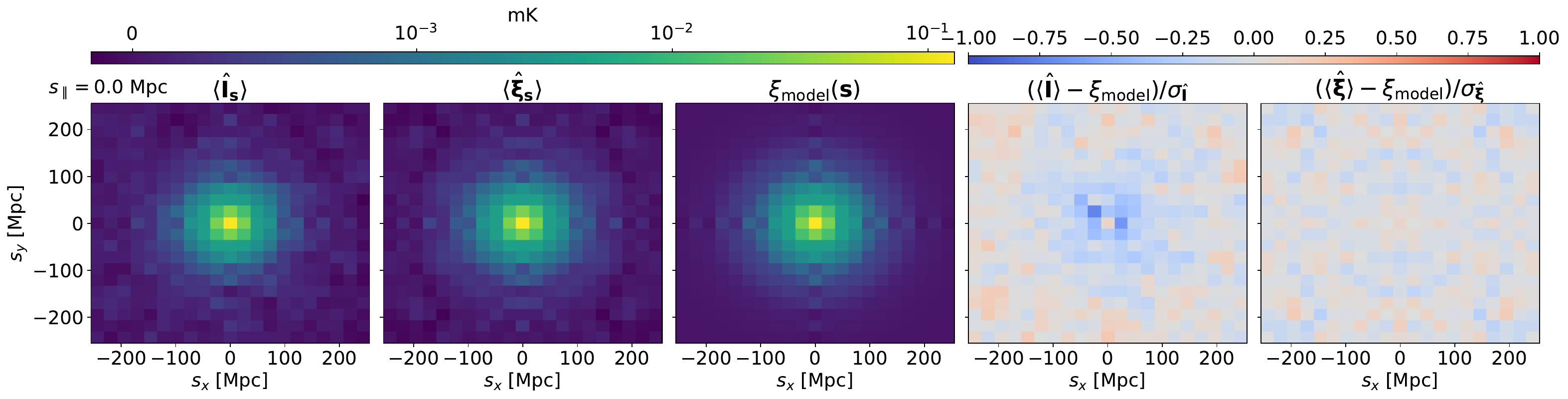}
    \includegraphics[width=\textwidth]{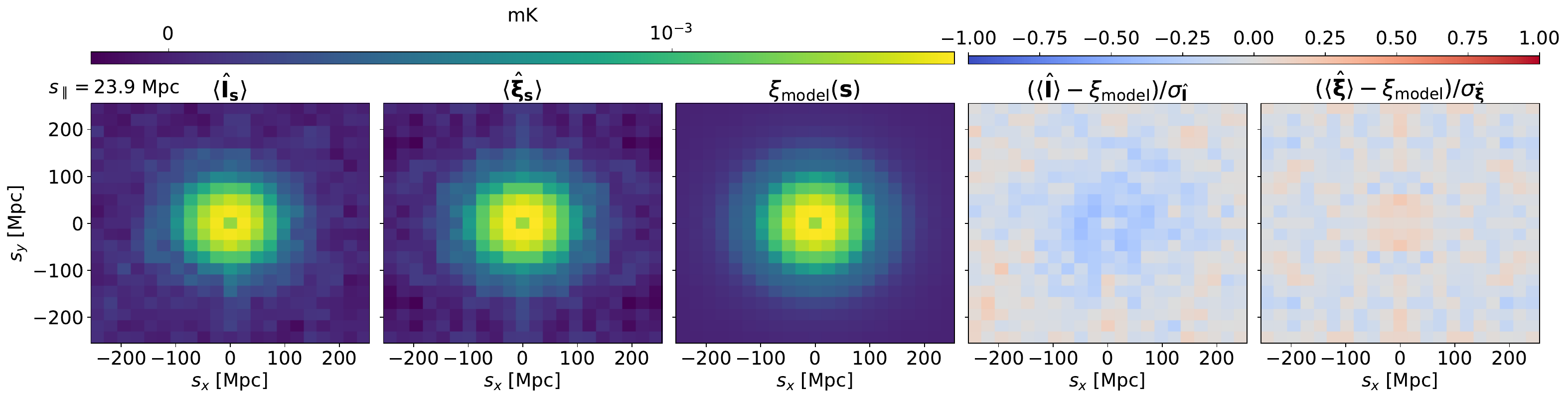}
    \includegraphics[width=\textwidth]{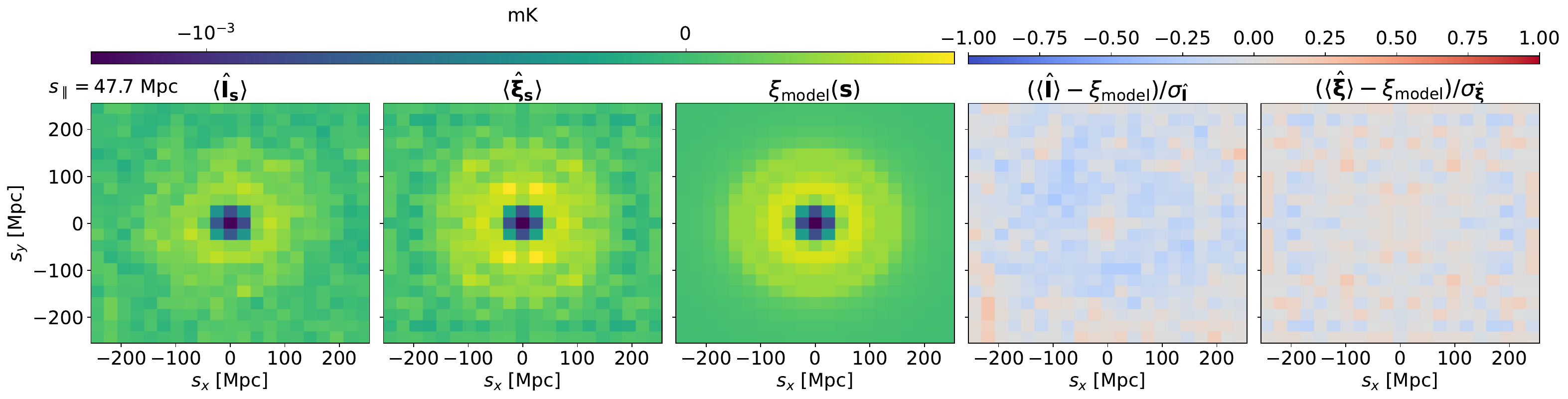}
    \caption{Validation of the equivalence between the stacking estimator and the two-point correlation function,
    using the pure \hi\ signal without any observational effects.
    From left to right, each panel shows the stacked signal in the transverse plane averaged over 100 realisations ($\langle \hat{\bm{I}}_{\bm{s}} \rangle$),
    the average of the estimated correlation function ($\langle\hat{\xi}_{\bm{s}}\rangle$),
    the input model ($\xi_{\rm model}(\bm{s})$),
    the difference between the stacked signal and the model over the standard deviation between the 100 realisations,
    and the difference between the estimated correlation function and the model over the standard deviation.
    For illustration, we show the results for three different separations along the line-of-sight,
    with $s_\parallel = 0$, $23.9$, and $47.7$ Mpc from top to bottom.
    }
    \label{fig:validate_ideal_angular}
\end{figure*}

In this section, we validate the stacking formalism by comparing the stacking estimator with the two-point statistics,
both the estimator of the cross-correlation function, \autoref{eq:stack_is_ift_ps}, and the theoretical model, \autoref{eq:stack_3d_with_pca}.
We assume ``Planck18'' cosmology \citep{2020A&A...641A...6P} throughout this section.

We use \textsc{meer21cm}\footnote{\url{https://meer21cm.readthedocs.io}} \citep{2026arXiv260701864C} to generate the simulations and perform the stacking analysis.
For illustration, we consider the case where the stacking is directly performed in rectangular boxes,
with lognormal simulations of \hi\ temperature fluctuations and galaxy catalogues, 
and ignore effects of map-making and gridding compensation.
We refer the reader to \citetaliasyear{2026arXiv260701864C} for more details
on the mock pipeline including these effects.

The simulation setup is summarised in \autoref{tab:sim_setup}.
\begin{table}
\caption{Setup for the mock stacking analysis.}
\label{tab:sim_setup}
\begin{ruledtabular}
\begin{tabular}{lc}
$z$ & 0.7 \\
Box size [Mpc] & $[1482.6,\,990.6,\,713.5]$ \\
Grid size & $[61,\,41,\,299]$ \\
Galaxy bias $b_g$ & 1.0 \\
$\bar{n}_{\rm gal}\,[{\rm Mpc}^{-3}]$ & $2\times10^{-5}$ \\
$\Omega_{\rm \hi}$ & $5\times10^{-4}$ \\
\hi\ bias $b_{\rm \hi}$ & 1.0 \\
$R_{\rm beam}$\,[Mpc] & 20.0 \\
\end{tabular}
\end{ruledtabular}
\end{table}
The model power spectrum is computed assuming a Kaiser RSD model,
and the details of the model power and lognormal simulation routine using the input power
can be found in \citetaliasyear{2026arXiv260701864C}.
The box size is chosen to correspond to a lightcone covering a redshift range of $0.6-0.8$,
over a sky area of $20\times 30$ square degrees.
The box resolution on the transverse plane is chosen to correspond to an angular resolution of
0.458 deg (\textsc{HEALPIX} \textsc{NSIDE}=128), consistent with the typical resolution of
single-dish \hi-intensity mapping surveys.
The resolution along the line of sight is chosen to correspond to
a frequency resolution of 0.2\,MHz, which is comparable to telescopes such as MeerKAT and SKA-Mid.
We simulate 100 random realisations of the mock stacking analysis.
The simulations are then used to perform estimation of the stacking signal, $\hat{\bm{I}}_{\bm{s}}$, and the correlation function, $\hat{\bm{\xi}}_{\bm{s}}$, for each realisation. The covariances of the estimators are then estimated from the realisations,
\begin{equation}
    {\mathbb{C}}_{\hat{\bm{I}}} = \frac{1}{N_{\rm sim}-1}\sum_{i=1}^{N_{\rm sim}} \left(\hat{\bm{I}}_i - \langle \hat{\bm{I}} \rangle\right) \left(\hat{\bm{I}}_i - \langle \hat{\bm{I}} \rangle\right)^\dagger,
\end{equation}
\begin{equation}
    {\mathbb{C}}_{\hat{\bm{\xi}}} = \frac{1}{N_{\rm sim}-1}\sum_{i=1}^{N_{\rm sim}} \left(\hat{\bm{\xi}}_i - \langle \hat{\bm{\xi}} \rangle\right) \left(\hat{\bm{\xi}}_i - \langle \hat{\bm{\xi}} \rangle\right)^\dagger,
\end{equation}
where $N_{\rm sim}$ is the number of realisations.
The square roots of the diagonal elements of the covariance matrices are then used to estimate the errors of the stacking signal $\sigma_{\hat{\bm{I}}}$ and the correlation function $\sigma_{\hat{\bm{\xi}}}$.

\subsection{Validation on the pure \hi\ signal}
\label{subsec:validate_ideal_case}

We first validate the most basic relation established in \secref{subsec:ideal_case},
that the stacked signal is the correlation function, i.e. the inverse Fourier transform of the power spectrum.
The simulated \hi\ signal is directly stacked onto the positions of the simulated galaxies,
with uniform weighting and without any observational effects.
In each realisation, we first measure the stacked signal using the conventional
formalism of averaging over the stacking cubelets, $\hat{\bm{I}}_{\bm{s}}$, as described in \autoref{eq:3D_stacking}.
We then measure the correlation function $\hat{\bm{\xi}}_{\bm{s}}$, by inverse Fourier transforming the estimated cross-power spectrum
on the 3D $\bm{k}$-space grid, $\hat{\bm{P}}_{\bm{k}}$.
To illustrate the comparison, we then choose slices of the measured signals at different values of $s_\parallel$,
and compare them with the input model in the transverse angular plane as shown in \autoref{fig:validate_ideal_angular}.

At $s_\parallel = 0$, where the signal is the strongest, both the stacked signal and the correlation function agree well with
the input model, $\xi_{\rm model}$, both in amplitude and in the scale-dependent angular structure as seen in the top row.
The consistency is maintained at other values of $s_\parallel$ as shown, at $s_\parallel = 23.9$\,Mpc, where
the correlation is much weaker, and further at $s_\parallel = 47.7$\,Mpc, where only a small anti-correlation
signal is expected due to integral constraints.
Compared against the input model,
both the measured stacked signal and the measured correlation function agree with the model within $1\sigma$ deviation
across all scales.

\begin{figure}
    \includegraphics[width=\columnwidth]{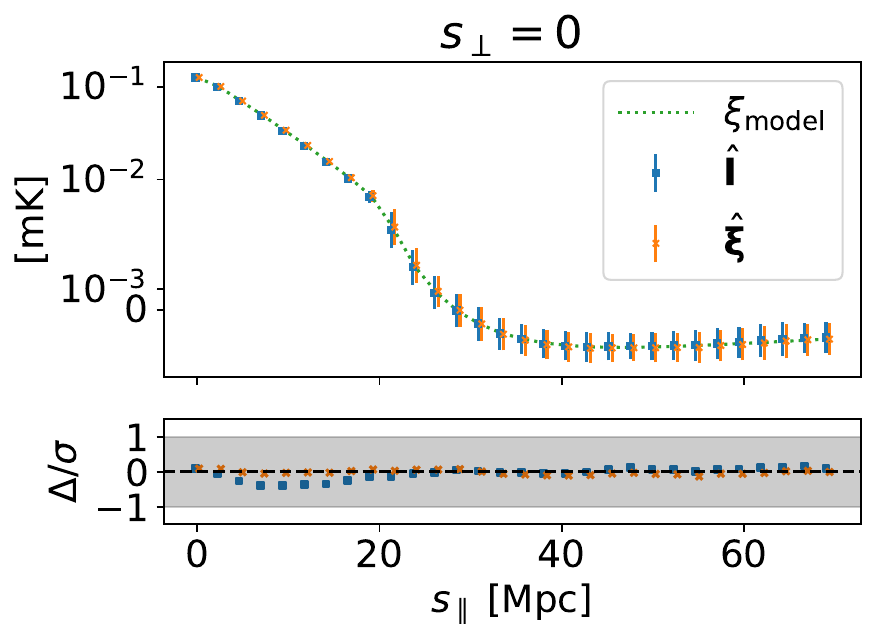}
    \caption{Validation of the equivalence between the stacking estimator and the two-point correlation function along the line-of-sight.
    The top panel shows the measured signals.
    The blue squares denote the estimated stacking signal ($\langle \hat{\bm{I}} \rangle$), averaged across 100 realisations.
    The orange crosses denote the average estimated correlation function ($\langle\hat{\bm{\xi}}\rangle$).
    The green dotted line denotes the input model ($\xi_{\rm model}$).
    The positions of the blue squares and orange crosses are shifted by -0.2 and +0.2 Mpc
    for better visibility.
    The bottom panel shows the differences between the measured signal
    and the model over the standard deviation.
    }
    \label{fig:validate_ideal_spectral}
\end{figure}

We also examine the equivalence between stacking and the two-point correlation function along the line-of-sight.
Since we have established in \autoref{fig:validate_ideal_angular} that the stacking signal agrees with the correlation function
in the transverse plane, we only compare $\bm{s}_\perp=0$ for different values of $s_\parallel$ for simplicity.
The results are shown in \autoref{fig:validate_ideal_spectral}.
The stacking signal and the estimated correlation function agree tightly with each other
across all line-of-sight scales.
Compared against the input model,
we see again that the measured signals agree with the model well within $1\sigma$ deviation.
Furthermore, the variances among the realisations
are within $\sim 1\%$ of each other for the two estimators{, emphasising the equivalence between the two estimators}.

The results in \autoref{fig:validate_ideal_angular} and \autoref{fig:validate_ideal_spectral}
convincingly demonstrate the equivalence between emission line stacking and the two-point correlation function.
The small numerical differences between the two estimators arise from a few technical details.
First, the stacked signal is estimated with a finite number of sources, which
introduces an additional Poisson noise on top of the galaxy overdensity field. 
Second, the derivation in \autoref{eq:stack_integral} turns
a discrete sum into a continuous integral, which assumes periodic boundary conditions
when relating the signal to the power spectrum.
Combined, these effects produce the negligible differences.

\subsection{Validation on mock observed \hi\ signal}
\label{subsec:validate_observed_case}

\begin{figure*}
    \includegraphics[width=\textwidth]{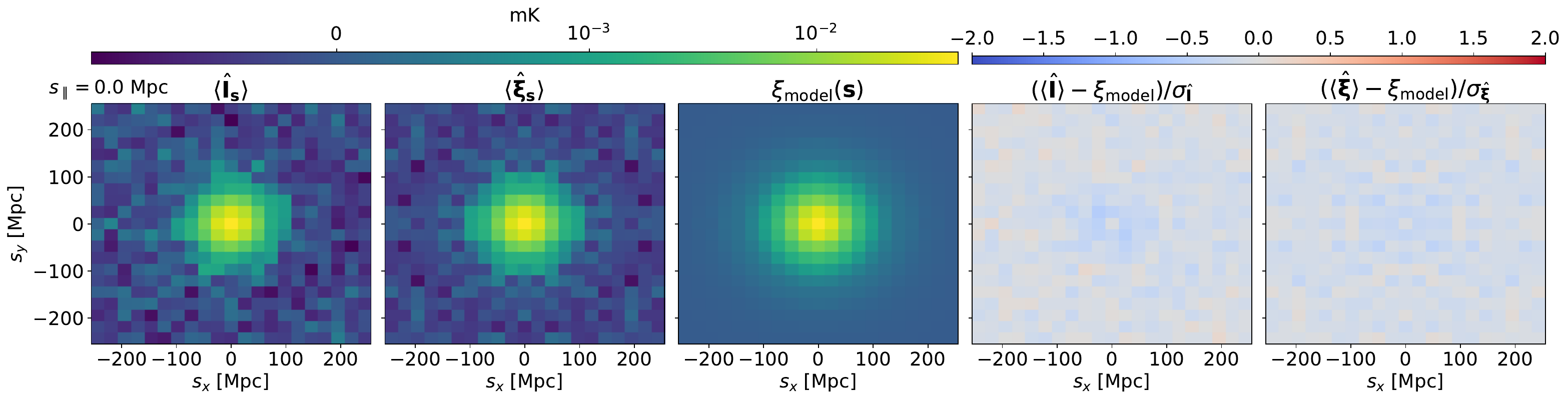}
    \includegraphics[width=\textwidth]{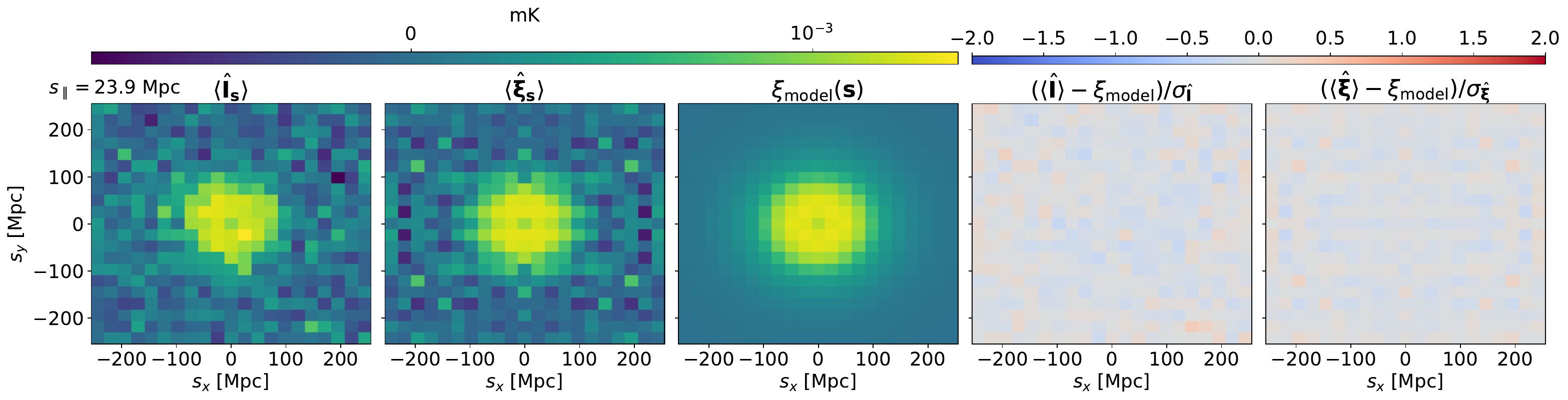}
    \includegraphics[width=\textwidth]{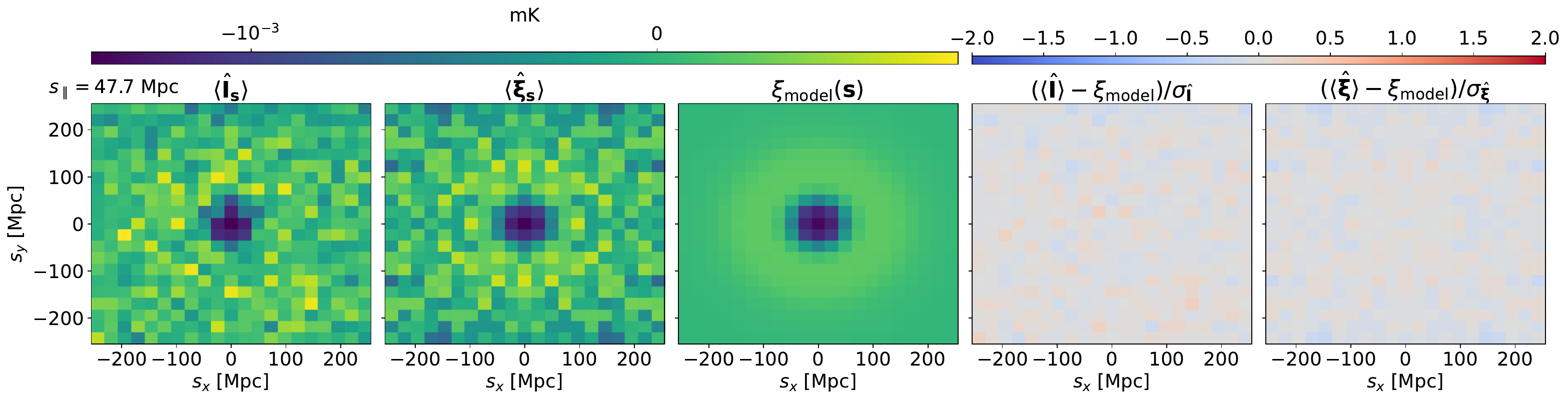}
    \caption{Validation similar to \autoref{fig:validate_ideal_angular},
    using the mock observed \hi\ signal with beam-smoothing, thermal noise, and foreground cleaning.
    }
    \label{fig:validate_observed_angular}
\end{figure*}

We then validate the stacking formalism of \autoref{eq:stack_3d_with_pca} in \secref{subsec:quadratic_general},
by including the effects of beam-smoothing, thermal noise, and foreground cleaning.
For simplicity, we stick to the same approximations and stack directly in the simulation box.
Foreground emission is added to the simulated temperature field,
by projecting a foreground template based on the Python Sky Model~\citep{2017MNRAS.469.2821T}.\footnote{The template added corresponds to the model [``d1'', ``s1'', ``f1'', ``a1'', ``c1''].}
We apply a Gaussian smoothing on the $x$-$y$ plane on the simulated total temperature field,
with a {standard deviation} of $R_{\rm beam} = 20$\,Mpc.
A thermal noise component is then added to the simulation cube,
with a standard deviation of $\sigma_{\rm noise} = 0.5$\,mK.
This number is chosen so that the thermal noise dominates over
the \hi\ signal at the voxel level, while being small enough
so that the variance at the estimator level does not dominate
over the effects of signal loss due to foreground cleaning.
For comparison, the MeerKLASS L-band deep-field survey has a noise level of $\sim1\,$mK in the map \cite{2025MNRAS.537.3632M}.

We then perform a PCA cleaning on the temperature field, removing $N_{\rm fg} = 5$ modes from the data.
It is not the focus of this work to simulate realistic foreground cleaning with instrument systematics,
as we are interested in demonstrating the mathematical equivalence.
Regardless of the fidelity of the foreground simulation,
the signal loss due to foreground cleaning can be modelled by the transfer function
as long as the cleaning is a matrix operation to the data vector,
as proved in Ref.~\cite{2025MNRAS.542L...1C}.
Correspondingly, the model power spectrum is modified by the beam attenuation factor,
and then multiplied by the transfer function before inverse Fourier transforming to the configuration space,
as described in \autoref{eq:stack_3d_with_pca}.

\begin{figure}
    \includegraphics[width=\columnwidth]{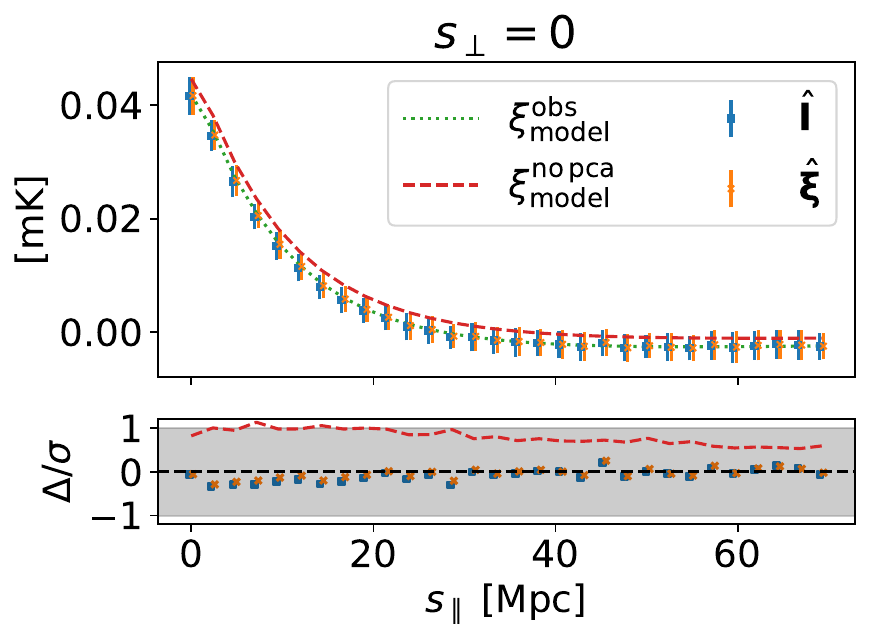}
    \caption{Validation similar to \autoref{fig:validate_ideal_spectral},
    using the mock observed \hi\ signal with beam-smoothing, thermal noise, and foreground cleaning.
    In the upper panel, the red dashed line denotes the model correlation function
    without the transfer function corrections ($\xi_{\rm model}^{\rm no\ pca}$), to demonstrate the effect of the foreground cleaning.
    }
    \label{fig:validate_observed_spectral}
\end{figure}

The results along the transverse direction are shown in \autoref{fig:validate_observed_angular}.
Compared against the ideal case,
the measured signals are smoothed near $|\bm{s}_\perp| \sim 0$ due to beam smoothing.
The estimated stacked signal again agrees well with the estimated correlation function,
and they are consistent with the model within $1\sigma$ deviation in the angular plane across different
values of $s_\parallel$.

The comparison along the line-of-sight is shown in \autoref{fig:validate_observed_spectral}.
The results show excellent agreement between the measured signals and the model correlation function,
with the model corrected by the transfer function as described in \autoref{eq:stack_3d_with_pca}.
Furthermore, the model without signal loss correction shows a significant mismatch with the measured signals,
with a consistent overestimation across all scales of $s_\parallel$.
For the level of thermal noise $\sigma_{\rm noise} = 0.5$\,mK,
the effect of signal loss creates a $\sim 1\sigma$ deviation
between the measured signals and the model.
Our results convincingly demonstrate the validity of 
the modelling framework for the stacked \hi\ signal
with observational effects.

\begin{figure}
    \includegraphics[width=\columnwidth]{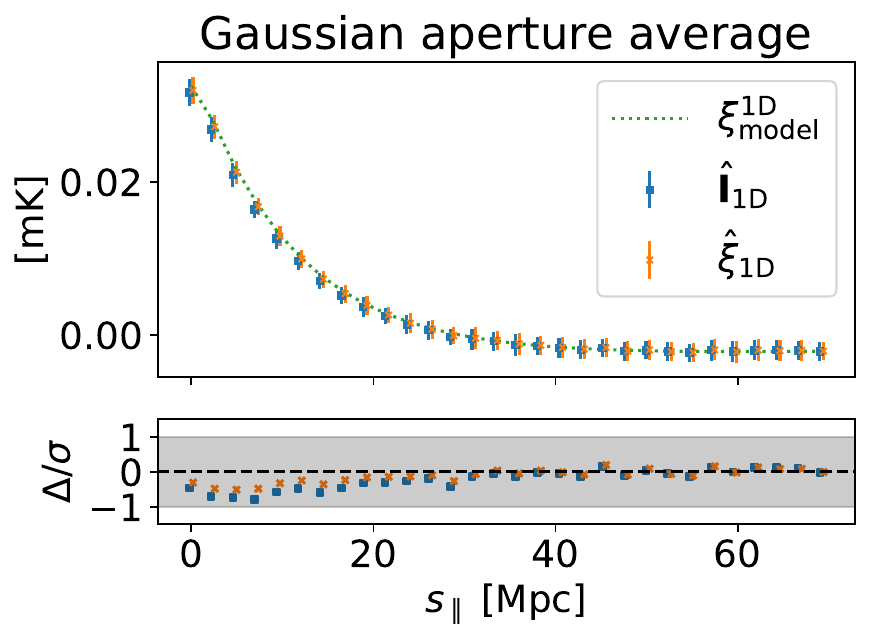}
    \caption{
        Validation of the equivalence between the 1D stacked spectrum ($\hat{\bm{I}}_{\rm 1D}$) and the 1D correlation function ($\hat{\bm{\xi}}_{\rm 1D}$).
        The dotted line denotes the model 1D correlation function ($\xi_{\rm model}^{\rm 1D}$) following \autoref{eq:1d_power}.
    }
    \label{fig:validate_1d}
\end{figure}

We proceed to validate the formalism of the transverse-integrated 1D stacking estimator
as defined in \autoref{eq:1D_stacking} and modelled according to \autoref{eq:1d_power}.
For illustration, we choose a Gaussian aperture with a width of $R_{\rm aperture} = 20$\,Mpc, {so that the window function featured in the selection function is given by}
\begin{equation}
    W_\perp(\bm{s}_\perp) = \exp\left(-\frac{|\bm{s}_\perp|^2}{2R_{\rm aperture}^2}\right).
\end{equation}
The results are shown in \autoref{fig:validate_1d}.
This test further validates the use of the correlation function inverse transforming power spectrum measurements for the stacking signal,
with deviations with respect to the model smaller than $1\sigma$. 
The deviation from the model grows at small distances $s_\parallel\lesssim 10$ Mpc, with the conventional stacking estimator showing larger discrepancies.

\begin{figure}
    \includegraphics[width=\columnwidth]{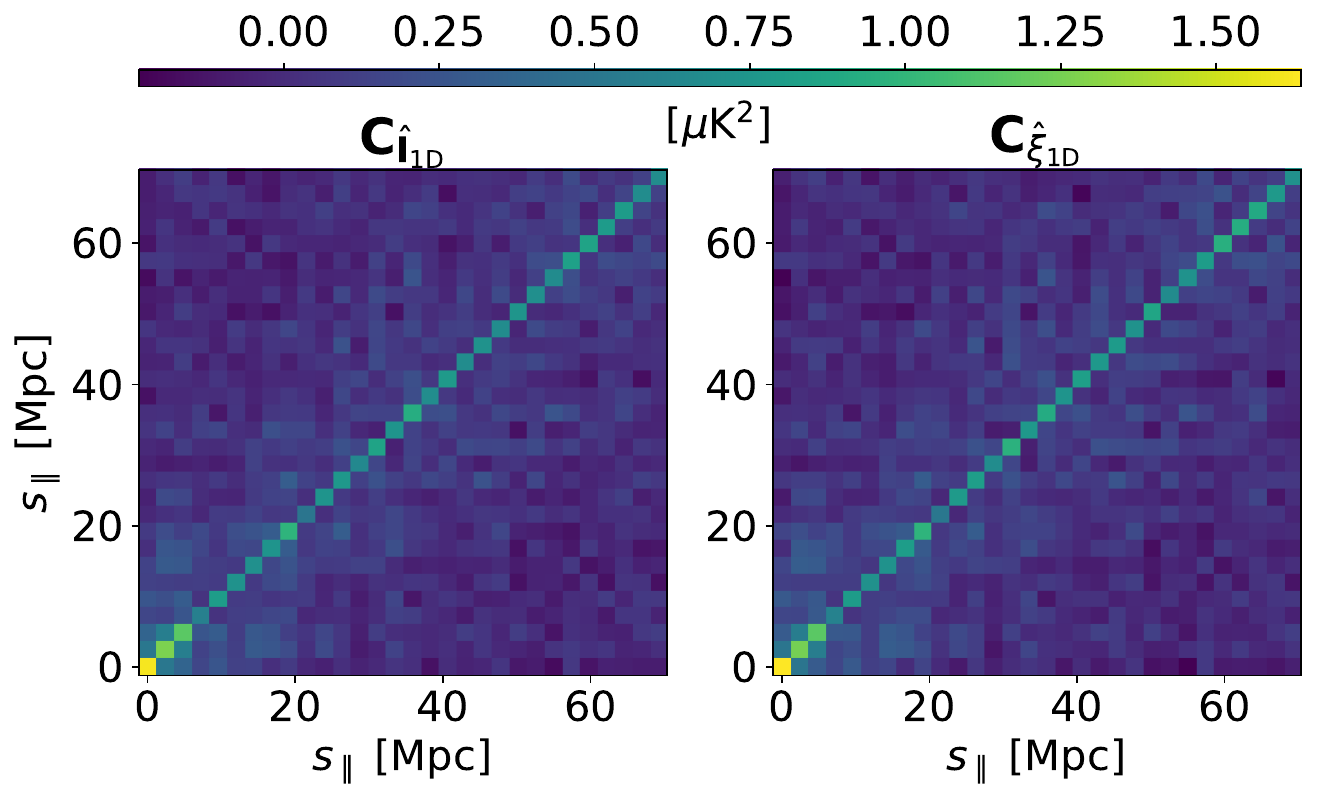}
    \caption{
        The mock-based covariance matrices for the stacked 1D spectrum (left) and the 1D correlation function (right).
    }
    \label{fig:validate_1d_covariance}
\end{figure}

Finally, the covariance of the stacked 1D spectrum can then be compared against the covariance of the correlation function.
The mock-based covariance matrices for the stacked 1D spectrum and the 1D correlation function are shown in \autoref{fig:validate_1d_covariance}.
The two covariance matrices are almost identical{, showing that the equivalence between the estimators extends to their covariances too. 
Therefore, there is no a priori gain from using one estimator or the other in terms of information content; we leave a further exploration of this matter for the stacking estimator without inverse transforming it back to configuration space for future work}.
To compare the covariance, we use the mean reduced chi-squared (see e.g., Ref.~\citep{2025JCAP...01..145R}),
\begin{equation}
    \chi^2_{\rm red} = \frac{1}{N_{\rm bins}}{\rm tr}\Big[ {\mathbb{C}}^{-1}_{\hat{\bm{I}}_{\rm 1D}} {\mathbb{C}}_{\hat{\bm{\xi}}_{\rm 1D}} \Big],
\end{equation}
where $N_{\rm bins}$ is the number of bins in $s_\parallel$.
For the covariance matrices shown, we obtain $\chi^2_{\rm red} = 1.078$.
As we have discussed in \secref{subsec:covariance}, in future data analysis,
we can utilise the covariance estimation of the power spectrum to construct the covariance of the stacked signal consistently.
Furthermore, it enables the modelling of the covariance between the stacked signal and the power spectrum measurements for joint analysis.

\section{Conclusions}
\label{sec:conclusion}
In this paper, we explicitly derive the equivalence between the emission line stacking measurement and the estimation of the two-point correlation function.
We show that measuring the 3D stacked signal around the positions of a catalogue of galaxies is exactly equal to
estimating the two-point cross-correlation function between the intensity map and the galaxy overdensity field. {In particular, we consider a correlation function obtained from the inverse Fourier transform of the measured power spectrum using a quadratic estimator.} 
The position vector from the galaxy to the voxel in the stacked cubelet is the position vector of the two-point correlation function in configuration space. We also demonstrate that in the shot-noise limit, the estimator returns to the standard picture where the stacked signal is simply an average of the emission line profile of the sources.

For a generic class of stacking procedures, {any linear} operation{, including} foreground cleaning,
map weighting, galaxy selection and aperture tapering{, among others,} can be factored into
the quadratic estimator of the two-point correlation function.
Therefore, the modelling of the stacked signal can be directly translated into the modelling of clustering cross-power spectrum, 
and the modelling of all the observational effects listed above is naturally embedded. 
This significantly simplifies the measurement and modelling of the stacked signal, speeding up the former ---especially in the case of a large number of galaxies--- and avoiding complicated implementation of observational effects in configuration space for the latter. 
Furthermore, although the interpretation of the stacked signal is significantly easier in configuration space, stacking measurements in Fourier space (i.e., without the inverse Fourier transform) entails other benefits, for instance a simpler covariance with less mode coupling.

The derivations are then validated against mock observed data.
As a proof of concept, we use simple lognormal simulations that include the effects of beam-smoothing, thermal noise, and foreground cleaning.
The results show excellent agreement between the measured 3D stacked signal and the estimated correlation function,
both of which agree with the input model well within $1\sigma$ deviation.
The agreement extends to both the angular and spectral directions, {and it is maintained when a selection function is applied to measure the 1D spectral stacking with a Gaussian aperture in the transverse direction, as well as to the covariance level. Hence, we expect the agreement to extend to any general stacking measurement.}

Our results have significant implications for the stacking analysis of LIM surveys as well as
conventional stacking analysis:
\begin{itemize}
    \item For LIM surveys, the stacked signal {usually} contains clustering signatures.
    Our results show that the modelling of the clustering contribution is simply the cross-correlation function. {Therefore, stacking analyses do not need to}
    rely on forward modelling, and this allows for fast Bayesian inference.
    \item The observational effects of the intensity mapping survey are fully accounted for in the modelling.
    For example, signal loss due to foreground cleaning can be modelled by the transfer function,
    which is part of the total window function of the correlation function quadratic estimator.
    \item Given a covariance estimation of the two-point cross-correlation function or cross-power spectrum,
    the covariance of the stacked signal can be constructed consistently.
    Parameter inference can be performed using the stacked signal and its covariance, optionally combined with the power spectrum measurements,
    without the need for additional efforts specifically for the stacking analysis.
    \item For large surveys such as MeerKLASS UHF in cross-correlation with DESI,
    the high number density of galaxies no longer presents a computational challenge,
    since the per-source stacking operation can be replaced by the estimation of the cross-correlation function. {This is especially true when different selection functions are applied, like in the case of aperture windows of different sizes.}
    \item For conventional stacking analysis, our method provides an interesting approach to modelling
    the effects of continuum subtraction and quantifying the measurement errors.
\end{itemize}

In conclusion, the formalism presented in this work provides a powerful tool for the analysis of stacking measurements in LIM surveys {and can be easily applied to any 3D stacking analysis of other observables}.
It provides accurate modelling of the signal and its covariance,
and significantly reduces the computational cost of the signal estimation as well as covariance calculation.
The method developed in this work will be applied to the analysis of the MeerKLASS UHF survey in cross-correlation with DESI,
in preparation for probing the \hi\ evolution of the Universe for MeerKLASS and future SKAO surveys.

\begin{acknowledgments}
ZC is funded by a UKRI Future Leaders Fellowship [grant MR/X005399/1; PI: Alkistis Pourtsidou]. 
JLB acknowledges funding from the project UC-LIME (PID2022-140670NA-I00), financed by MCIN/AEI/10.13039/501100011033/FEDER, UE, and from the project DiffWeb (PID2025-173927NB-I00), financed by MICIU/AEI/10.13039/501100011033. The Instituto de Física de Cantabria (IFCA) has been accredited as Unidad de Excelencia María de Maeztu, a distinction awarded by the Spanish State Research Agency for the period 01/06/2026 - 31/05/2030.
\end{acknowledgments}


\appendix

\section{Stacking for general selection functions}
\label{apdx:general}
Depending on the specific observable there are different optimal choices for the selection function. 
In the specific case of LIM, selection functions proportional to Dirac deltas have been very useful in the regime of high signal-to-noise ratio.
They can be used to probe the emission profile around galaxies (see e.g., Refs.~\cite{2022ApJ...929...90L, 2022ApJ...934L..26L}),
or to include the clustering term and check for systematics~\cite{2025ApJS..279...19C}.
On the other hand, in the low signal-to-noise ratio regime, using a selection function following an assumed line and angular profile of the signal can help to maximise the detection significance, turning the whole stacking signal into a single scalar~\cite{2026ApJ..1005....8M}. 
Some other observables, like the kSZ effect, may benefit from compensated aperture photometry, which reduces the stacking to a series of scalar values for each aperture value used in the selection function, to minimise the impact of the long wavelength modes and impact of the clustering (see e.g., Ref.~\cite{2025arXiv251214625H}). 
All these cases, even for directional stacking which targets dipole signals, can be reframed in terms of the power spectrum or correlation function~\cite{2026arXiv260515947H} using the quadratic estimator described in this work.

In this appendix we consider the most general stacking (neglecting any directionality, which can be implemented trivially by finding the analogy from Ref.~\cite{2026arXiv260515947H}) and derive the final expression for its signal including all observational effects. 
This is then a generalisation of \secref{subsec:projection} for an arbitrary projection. 
Consider a case similar to the aperture photometry for different aperture values, in which the 3D cubelet is projected using a selection function $\bm{S}_{\bm{s}}^{\bm{d}}$ to a series of scalar values indexed by $\bm{d}$. 
The signal for each of these stacked projections is given by
\begin{equation}
    \hat{\bm{I}}_{\bm{d}}=\frac{1}{Q_{\bm{d}}}\sum_{\bm{s}}\bm{S}_{\bm{s}}^{\bm{d}}\,\hat{\bm{I}}_{\bm{s}}\,,
\end{equation}
{where $Q_{\bm{d}}\equiv\sum_{\bm{s}}\bm{S}_{\bm{s}}^{\bm{d}}$. If we expand the estimator for the 3D cubelet stacking using \autoref{eq:stack_3d_with_pca}, we find}
\begin{equation}
\begin{split}
     \hat{\bm{I}}_{\bm{d}} & =\frac{1}{Q_{\bm{d}}Q_0'}\sum_{\bm{sk}}\bm{S}_{\bm{s}}^{\bm{d}}\,\mathbb{F}_{\bm{sk}}^{-1}\,P_{\bm k}\,\bar{Q}_{\bm{k}}\,\mathcal{T}(k_\perp) \\
    &  = \frac{1}{Q_{\bm{d}}Q_0'}\sum_{\bm{skk'}}\,\mathbb{F}_{\bm{k}'\bm{s}}\,\tilde{\bm{S}}_{\bm{k}'}^{\bm{d}}\,\mathbb{F}_{\bm{sk}}^{-1}\,P_{\bm k}\,\bar{Q}_{\bm{k}}\,\mathcal{T}(k_\parallel) \\
    & = \sum_{\bm{k}}\,\tilde{\bm{S}}_{\bm{k}}^{\bm{d}}\,P_{\bm k}\,\bar{Q}_{\bm{k}}\,\mathcal{T}(k_\parallel)\,.
\end{split}
\end{equation}
{This expression is equivalent to the zero-lag correlation function of the observed galaxy number density field and the intensity field convolved with $\bm{S}_{\bm{s}}^{\bm{d}}$ (see \autoref{eq:3D_stacking_observed_WLs}) or, similarly, to the inverse Fourier transform of the observed power spectrum, but using $\tilde{\bm{S}}_{\bm{k}}^{\bm{d}}$ instead of the operator $\mathbb{F}_{\bm{sk}}^{-1}$, as pointed out in Ref.~\cite{2025arXiv251214625H} for ideal cases. }

This result is very powerful, because it allows for computing \textit{any} kind of stacking from a \textit{single} power spectrum measurement, significantly speeding up the calculation of the estimator from the observations. 
On top of it, it provides a clearer picture of the relative information encoded when using different selection functions for the stacking and enables the study of correlations among them for potential combinations of different stacks in order to increase the robustness of the measurement.

\bibliography{example}
\bibliographystyle{utcaps}

\end{document}